\documentclass[aip,cha,preprint,numerical,superscriptaddress]{revtex4-1}
\usepackage{graphicx}
\usepackage{epstopdf}
\usepackage{amsfonts}
\usepackage{amsmath}
\usepackage{bm}
\usepackage{enumerate}
\usepackage{color}
\begin{document}

\title{Stochastic bursting in unidirectionally delay-coupled noisy excitable systems}
\author{Chunming Zheng}
\affiliation{Institute for Physics and Astronomy, University of Potsdam, Karl-Liebknecht-Strasse 24/25, 14476 Potsdam-Golm, Germany}
\author{Arkady Pikovsky}
\affiliation{Institute for Physics and Astronomy, University of Potsdam, Karl-Liebknecht-Strasse 24/25, 14476 Potsdam-Golm, Germany}
\affiliation{Control Theory Department, Institute of Information Technologies, 
Mathematics and Mechanics, Lobachevsky University Nizhny Novgorod,603950, Russia}
\date{\today}

\begin{abstract}
We show that \emph{stochastic bursting} is observed in a ring of unidirectional 
delay-coupled noisy excitable systems, thanks to the combinational action 
of time-delayed coupling and noise. 
Under the approximation of timescale separation, i.e., when 
the time delays in each connection  are much 
larger than the characteristic duration of the spikes, the observed rather 
coherent spike pattern can be 
described by  an idealized  coupled point process with a leader-follower 
relationship. We derive analytically the statistics of the spikes in each unit, 
the pairwise correlations 
between any two units, and the spectrum of the total output from the network.
Theory is in a good agreement with the simulations with a network of theta-neurons.
\end{abstract}


\maketitle
\begin{quotation}
Excitable systems are basically in a resting state, but can generate a strong output
under a small but finite forcing. A prominent and very important example in neuroscience is
a neuron, which produces a spike when the input forcing is strong enough.
Under noisy action, an excitable neuron produces a sequence of random spikes.
In this paper we show, that with an additional time-delayed coupling, a network
of noisy excitable neurons can produce rather coherent bursts - sequences of spikes
separated by fixed delay times; the number of spikes in a burst is random. We construct
a point process model for this stochastic bursting, and derive analytically
the properties of the interspike intervals, and of the correlations and the spectra.
\end{quotation}

\section{Introduction}
Processes in delay-coupled nonlinear elements have been attracting a lot of attention, 
in oscillators (or neurons) \cite{yeung1999time,rosenblum2004controlling,perlikowski2010periodic}, 
laser dynamics \cite{soriano2013complex}, and complex 
networks \cite{d2008synchronization,flunkert2010synchronizing}. 
While for deterministic oscillators the major interest is in synchronization phenomena,
for noise-induced processes time-delayed coupling is known to effect strongly
the coherence and the correlation properties of the processes. 

There are two basic models of
noise-induced processes: noise-induced switchings between two stable states (resulting
in a telegraph-type stochastic process), and noise-induced pulses in an excitable system.
For the former situation, the cross-correlations between bistable 
units were investigated in unidirectional delay-coupled networks~\cite{kimizuka2009stochastic}, 
by extending the theory of a lump bistable noisy model with a delayed feedback~\cite{tsimring2001noise}.
In another approaches, one studies delay-coupled systems with noise within the 
mean field approximation framework~\cite{huber2003dynamics,hasegawa2004augmented}, 
focusing on the evolution of the mean and the variance of the order parameter while 
igonoring the correlation of different units. 
A full understanding of situations with more complex coupling topology, 
e.g. for the all-to-all coupling\cite{kimizuka2010stochastic}, is still an 
ongoing subject of research. 

Noise-induced pulses in an excitable unit is one of the basic models
in neuroscience. In this context, delayed feedbacks and couplings
are very natural due to a finite time of pulse propagation in connecting synapses. 
In our previous paper \cite{zheng2018delay}, a novel delay-induced spiking pattern 
which we called \emph{stochastic bursting}, was observed  in a single noisy excitable 
system with a time-delayed feedback (in the context of neuroscience this corresponds to an
autapse with a finite propagation time). This stochastic bursting can be 
characterized as a random sequence of quasiregular patches, with a pronounced
peak in the spectrum at the frequency corresponding to the delay time.

In this paper we extend the theory~\cite{zheng2018delay} to the case 
of two mutually coupled excitable units, and further
to a simple, but widely used,  topology of an unidirectional 
delay-coupled network. After outlining the main features and approximations
behind the theory of one unit, we describe \emph{stochastic bursting} in two coupled
units in details. The generalization to a chain of neurons will be then straightforward.

\section{Basic model and properties of one unit}
\begin{figure}
	\centering
	\includegraphics[width=0.3\textwidth]{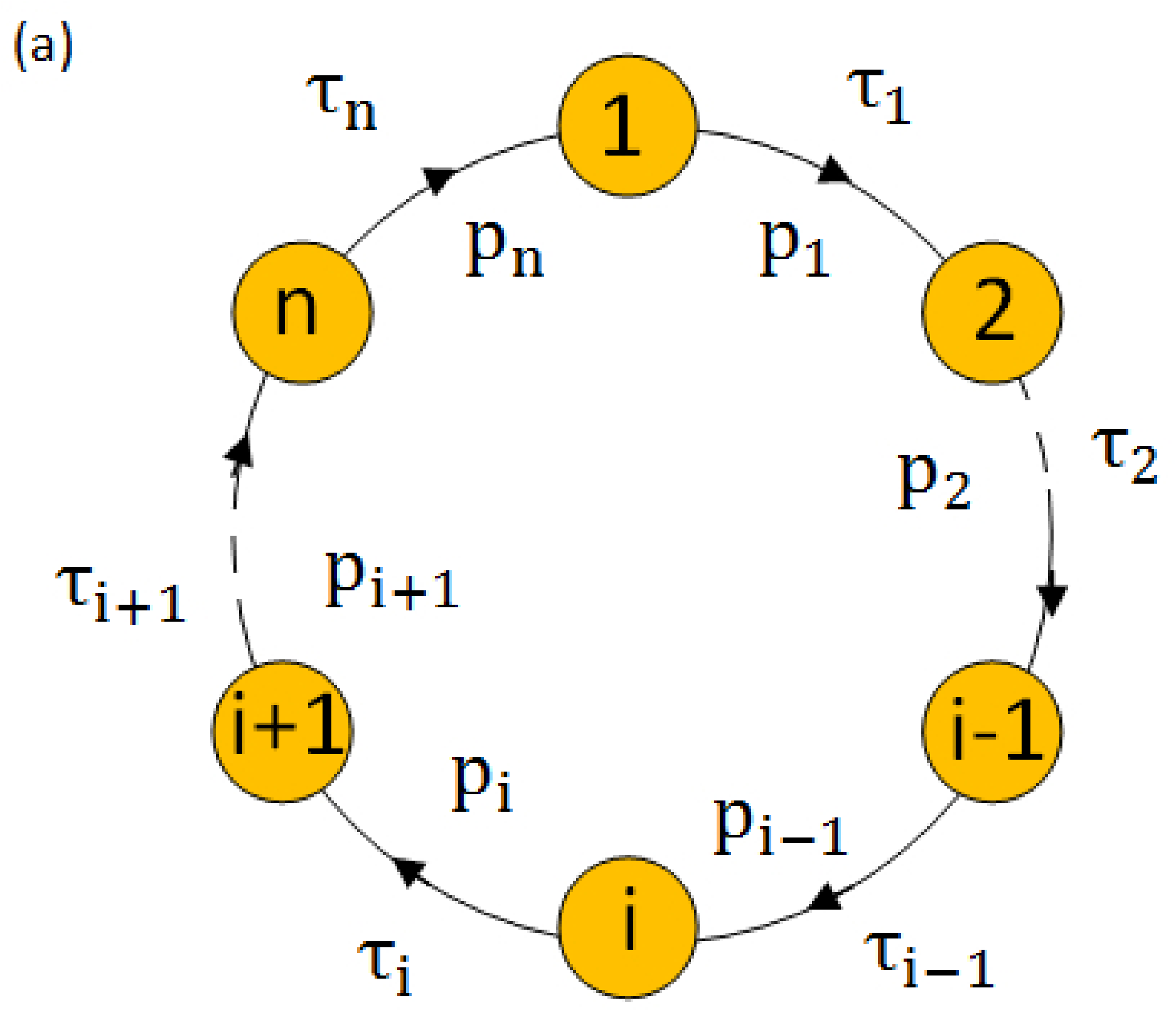}\\
	\includegraphics[width=0.8\textwidth]{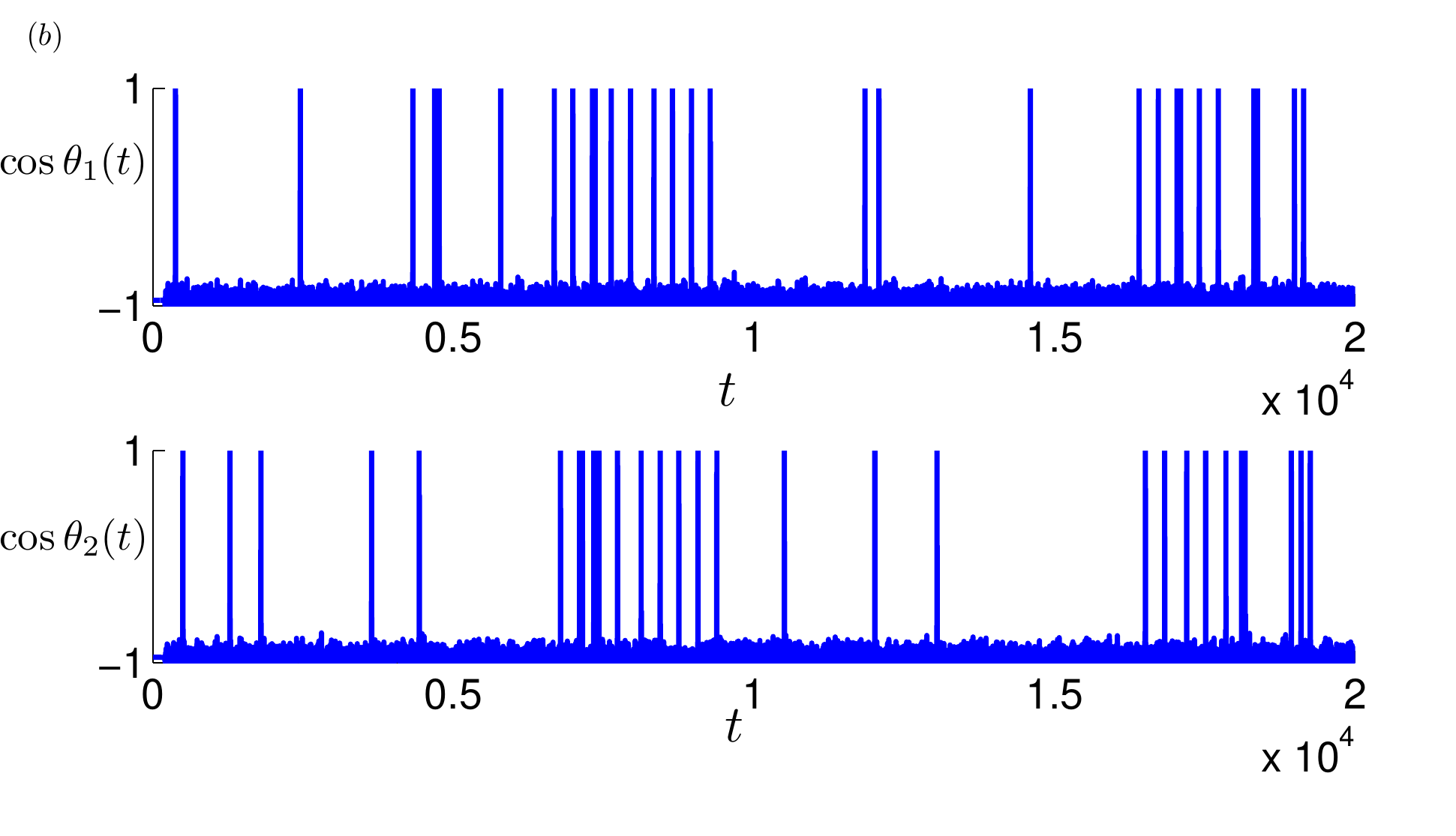}
	\caption{Panel (a): Schematic description of a ring of unidirectional delay-coupled 
	noisy systems, where a spike in neuron $i$ 
	 induces a spike to neuron $i+1$ after delay time $\tau_{i}$ with probability $p_{i}$. 
	Panel (b): The spike trains in a two-neuron network,
	obtained in direct simulations of Eq.~(\ref{Eq:model}). Values of parameters:
	 $a=0.95$, $D=0.005$, $\hat{\tau}_{1}=100$, $\hat{\tau}_{2}=200$, 
	and $\epsilon=0.14$. }
	\label{fig:spike}
\end{figure}

We consider in this paper a network of unidirectionally coupled units, topology of which
is illustrated in Fig.~\ref{fig:spike}a. The units are generally different, and 
the propagation times for the interactions are also different. Each unit is described 
with a prototypic model for an excitable system, a 
noisy theta-neuron~\cite{ermentrout1986parabolic} (or, in
other contexts, called active rotator~\cite{shinomoto1986phase}):
\begin{equation}\label{Eq:model}
\dot{\theta_i}=a_i+\cos\theta_i+
\epsilon_i(a_{i-1}+\cos\theta_{i-1}(t-\hat{\tau}_{i-1}))+\sqrt{D_i}\xi_i(t).
\end{equation}
Variable $\theta$ is defined on a circle $0\leq \theta<2\pi$.
Here parameters $a_i$ define the excitability property of the neurons. For $a_i<1$,
there is a stable and an unstable steady states for an isolated unforced unit, and 
these states collide in a SNIPER bifurcation at ${a_i=1}$. Thus, 
close to this threshold, the
unit is excitable: a small noise or a small force may induce a spike (nearly
$2\pi$-rotation of $\theta$ back to the stable state on the circle). Parameter 
$D_i$ 
describes intensity of the Gaussian white noises $\xi_i(t)$, with 
$\langle \xi_i(t)\rangle=0$,   $\langle
\xi_i(t)\xi_j(t^{'})\rangle=2\delta_{ij}\delta(t-t^{'})$. Finally, parameters
$\epsilon_i$ describe the 
strengths of delayed coupling. The coupling force,
amplitude of which is $\epsilon_i$, is chosen to vanish in the
steady state of a driving unit. The 
forcing term on the r.h.s. of \eqref{Eq:model} produced by 
one spike can be represented as
\begin{equation}
H(t)=a+\cos\Theta_{sp}(t),
\label{Eq:forc}
\end{equation}
with $\Theta_{sp}(t)$ being the deterministic trajectory
connecting the unstable point (the threshold) with the stable one: 
\begin{equation}
\Theta_{sp}(t)=2\arctan \left(\sqrt{\frac{1+a}{1-a}}
\tanh\left(\frac{\sqrt{1-a^2}}{2}(t-t_{0})\right)\right)\;.
\label{Eq:spike}
\end{equation}

\section{Network dynamics and the point process model}

To describe qualitatively the dynamics in the network,
we start with a non-coupled unit. For a small noise, it produces
independent spikes,  constituting a Poisson process
with rate $\lambda$. Calculation of this rate is a standard task. One
formulates the Fokker-Planck equation 
for the evolution of the probability density of a noisy unit 
obeying Eq.~(\ref{Eq:model}):
\begin{equation}
\frac{\partial P(\theta,t)}{\partial t}=-\frac{\partial }{\partial\theta}\left[(a+\cos\theta)
P(\theta,t)\right]+D\frac{\partial^{2} P(\theta,t)}{\partial\theta^{2}}.
\label{Eq:fpe}
\end{equation}
The stationary solution of \eqref{Eq:fpe} is
\begin{equation}
P_{st}(\theta)=C\int_{\theta}^{\theta+2\pi}\frac{d\psi}{D}e^{-\int_{\theta}^{\psi}\frac{a+\cos\varphi}{D}d\varphi}.
\label{Eq:stationary_withoutdelay}
\end{equation}
Here $C$ is a normalization constant. Then the probability current yields the 
rate of spontaneous spike excitations: 
\begin{equation}
\lambda=C\left(1-e^{-\int_{0}^{2\pi}\frac{a+\cos\theta}{D}d\theta}\right). \label{Eq:current}
\end{equation}

With coupling,  i.e. with $\epsilon\neq 0$, spikes of neuron $i-1$ produce, with
a delay, a kick to its next neighbor $i$. Such a kick facilitates excitation,
so that it   will cause a pulse in neuron $i$,
as a combinational effect of forcing and noise, with probability $p_i$.
The timing of the induced spike is slightly shifted to the forcing,
we will denote this shift $\bar\tau$ and define a new effective 
delay $\tau_i=\hat{\tau}_i+\bar \tau$ 
(we will mainly consider the shift $\bar\tau$ as a fixed one, but will
briefly discuss the effect of fluctuations of this quantity in section~\ref{sec:concl}.)
Hence, to be more general,  units are described by different quantities $\lambda_i$ 
(rates with which they ``spontaneously''
produce spikes due to noise), $p_i$ (the probability with which a spike is induced by
the incoming force), and $\tau_i$ (time delay in forcing).

This allows us to describe the activity on the network as a point process,
in which we neglect the durations of the spikes (approximate them as delta-functions),
compared to the delay times $\tau$ and the inverse rate $\lambda^{-1}$.
This is well justified for mammal brains, where the characteristic duration
of a spike is $\sim 1ms$, while the delay time and the characteristic 
time
interval between noise-induced spikes are of order 
$\sim 100 ms$~\cite{Swadlow:2012,Longtin:2013}.
The spike occurred at time $t$ in neuron $i$, will produce a kick on 
neuron $i+1$ at time $t+\hat{\tau}_{i}$, and will generate a spike
in neuron $i+1$ at 
time $t+\tau_{i}$ with probability $p_{i}$, as depicted 
schematically in Fig.~\ref{fig:spike} (a). 

Throughout 
the paper, in numerical illustrations we use parameters $a=0.95$, $D=0.005$.
The small additional delay is $\bar\tau \approx 7$,
it is much smaller than the characteristic delay times we use $\hat\tau\gtrsim 100$
and the inverse of the spontaneous rate $\lambda^{-1}\approx 1506$. 
For these parameters
of the neuron, we use the coupling strength $\epsilon=0.14$, for which
the probability to induce a spike by the forcing is $p=0.53$ (for details
of the calculation of this probability see Ref.~\onlinecite{zheng2018delay}). 
Empirically,
this probability can be determined in simulations of one unit
with a delayed self-feedback. One calculates the numbers of spikes, during
a large time interval, in dependence on the coupling strength $N(\epsilon)$. Then,
\begin{equation}\label{Eq:p_simulation}
p(\epsilon)=\frac{N(\epsilon) -N(0)}{ N(\epsilon)}\;.
\end{equation}

\begin{figure}
	\centering
    \includegraphics[width=\textwidth]{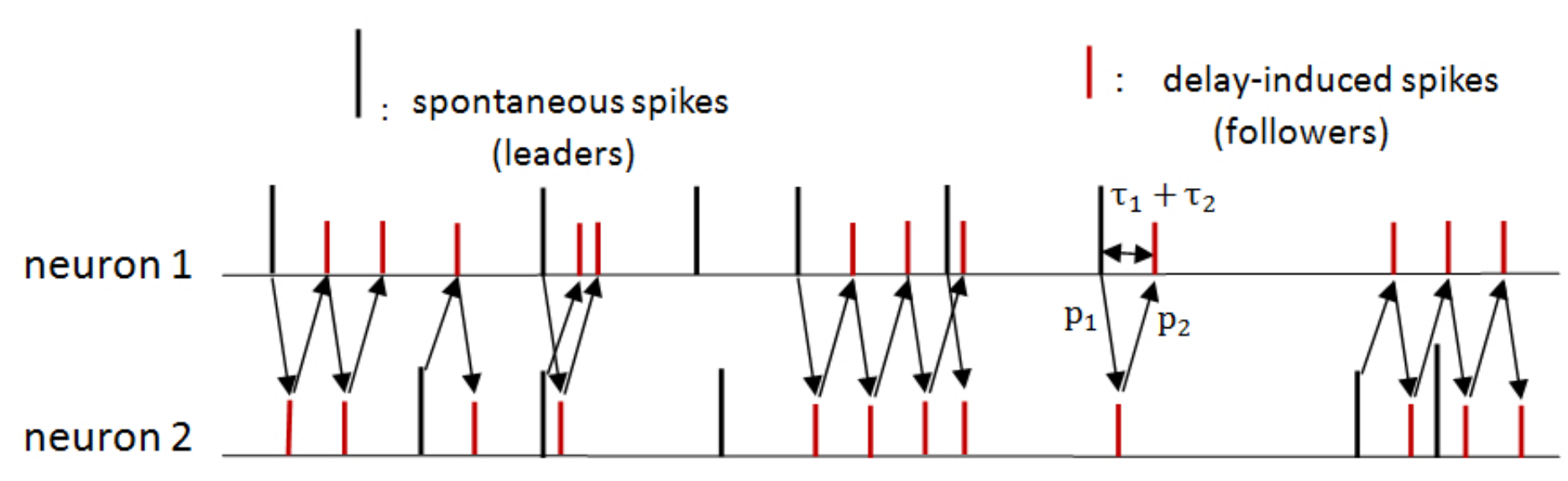}
	\caption{Schematic description of the point process with the leader-follower 
	relationship for two neurons. 
	A leader with a random number of its followers 
	form a burst, and the followers could be in both neurons. 
	The leaders appear in both neurons.}
	\label{fig:pointproc1}
\end{figure}

\section{Two coupled units}
\subsection{Statistics of interspike intervals}
It is instructive to start with the case where there are only two neurons in the ring, 
i.e. $n=2$, and then to extend the theory to a more general case with arbitrary $n>2$. 
When $n=2$, 
the two delay-coupled neurons are denoted as $i$ and $j$ ($i,j$ could be $1$ or $2$). 
We simulate Eq.~(\ref{Eq:model}) and obtain spike trains with bursts of each neuron as 
shown in Fig.~\ref{fig:spike}(b). As outlined above, an idealized point process 
can be constructed to describe the bursting phenomenon, as 
illustrated in Fig.~\ref{fig:pointproc1}. Spontaneously generated spikes 
we denote as leaders. Each leader induces a finite set of
followers (induced spikes), and together with them constitutes a coherent burst. In a burst,
spikes in the two neurons appear alternatively with time intervals $\tau_1$ and $\tau_2$.
The number of spikes in a burst is random.
Noteworthy, similar to the case of one neuron with delayed feedback~\cite{zheng2018delay}, the 
bursts can overlap; hence the analysis of the ISI (inter-spike intervals) distribution 
of the spike train in each unit is nontrivial.
Compared to the single neuron case~\cite{zheng2018delay}, here the leaders 
in each neuron will have random followers in both neurons, as explained schematically 
in Fig.~\ref{fig:pointproc1}. 

First, we determine the overall rate of the spikes in each unit. The probability for a spike
in unit $i$ to induce a follower in the same unit is $p_ip_j$. Thus, the
probability for a leader to have exactly $m$ followers in 
the same unit is $(p_ip_j)^m(1-p_ip_j)$.
The average number of followers in the same unit is $
\sum_m m(p_ip_j)^m(1-p_ip_j)= (p_i p_j)(1-p_ip_j)^{-1}$.
The total average number of spikes in a burst is $1+(p_i p_j)(1-p_ip_j)^{-1}=(1-p_ip_j)^{-1}$
Therefore, the total rate of spikes initiated in unit $i$ is $\lambda_i(1-p_ip_j)^{-1}$. 
For the spikes in unit $i$, initiated by a leader in unit $j$, we have first to find the rate
of the first followers in unit $i$, which is $\lambda_j p_j$; the total rate of these
spikes is thus $\lambda_jp_j (1-p_ip_j)^{-1}$. Summing, we obtain the total rate
of spikes $\mu_i$ as
\begin{equation}
\begin{aligned}
\mu_{i}&=\lim\limits_{m\rightarrow\infty}(\lambda_{i}+\lambda_{j}p_{j})[1+p_{i}p_{j}+(p_{i}p_{j})^2+...+(p_{i}p_{j})^m]=\frac{\lambda_{i}+\lambda_{j}p_{j}}{1-p_{i}p_{j}}.
\end{aligned}\label{Eq:spikerate_2neurons}
\end{equation}

To derive the statistics of the ISI, we assume that in 
one of the units there is a spike at time $t$, and the next spike at
time $t'>t$, so that the inter-spike interval is $T=t'-t$. Three 
different cases should be distinguished, 
namely, $T>\tau_{i}+\tau_{j}, T=\tau_{i}+\tau_{j}$, and $T<\tau_{i}+\tau_{j}$. 
 If $T<\tau_{i}+\tau_{j}$, the spikes at $t$ and $t'$ 
can be either spontaneous (leader) or delay-induced ones, but in the latter case they belong
to different bursts, so they are independent. Therefore,   the survival function, 
i.e., the probability that there is no spike in $(t, t')$, is
determined by the full rate $\mu_{i}$ from~\eqref{Eq:spikerate_2neurons}:  
$S_{i}(T)=\exp(-\mu_{i}T)$.

In contradistinction, for the case $T>\tau_{i}+\tau_{j}$, the spike at time $t'$ in 
neuron $i$ can only be from a spontaneous one (leader) in neuron $i$ itself, 
or the first induced spike (with probability $p_{j}$ from neuron $j$). 
These events are independent on the occurrence of a spike at time $t$, and have
the total rate $\lambda_{i}+\lambda_{j}p_{j}$.
The probability that there is no any
spike in $(t, t')$ in neuron $i$ is  the product of three terms: 
the probability to have  no 
spikes in the interval $(t, t+\tau_{i}+\tau_{j}]$ with the survival function 
$S_{\tau b}=\exp(-\mu_{i}(\tau_{i}+\tau_{j}))$, the probability
$(1-p_{i}p_{j})$ not to have a follower for the spike at $t$, 
and the probability
to have no spike in the interval $(t+\tau_{i}+\tau_{j}, t')$, where only the spontaneous 
total rate $\lambda_{i}+\lambda_{j}p_{j}$ applies with the survival function  
$S_{\tau a}=\exp(-(\lambda_{i}+\lambda_{j}p_{j})(T-\tau_{i}-\tau_{j}))$.
Thus, the survival function for the case $T>\tau_{i}+\tau_{j}$ is 
\begin{equation}
\begin{aligned}
S_{i}(T)&=S_{\tau b}(1-p_{i}p_{j})S_{\tau a}=
(1-p_{i}p_{j})e^{-\mu_{i}(\tau_{i}+\tau_{j})-(\lambda_{i}+\lambda_{j}p_{j})(T-\tau_{i}-\tau_{j})}.
\end{aligned}
\end{equation}

\begin{figure}
	\centering
	\includegraphics[width=\textwidth]{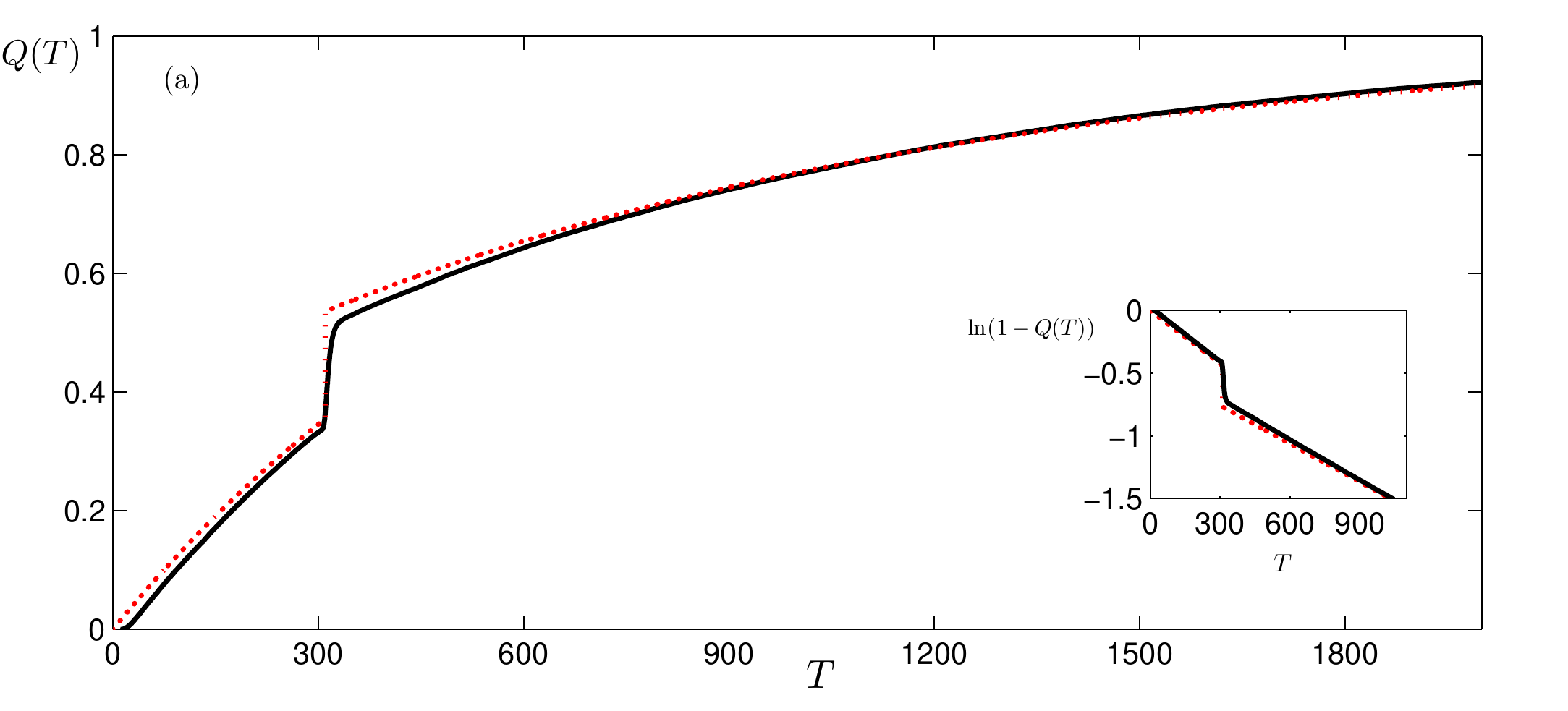}
	\includegraphics[width=\textwidth]{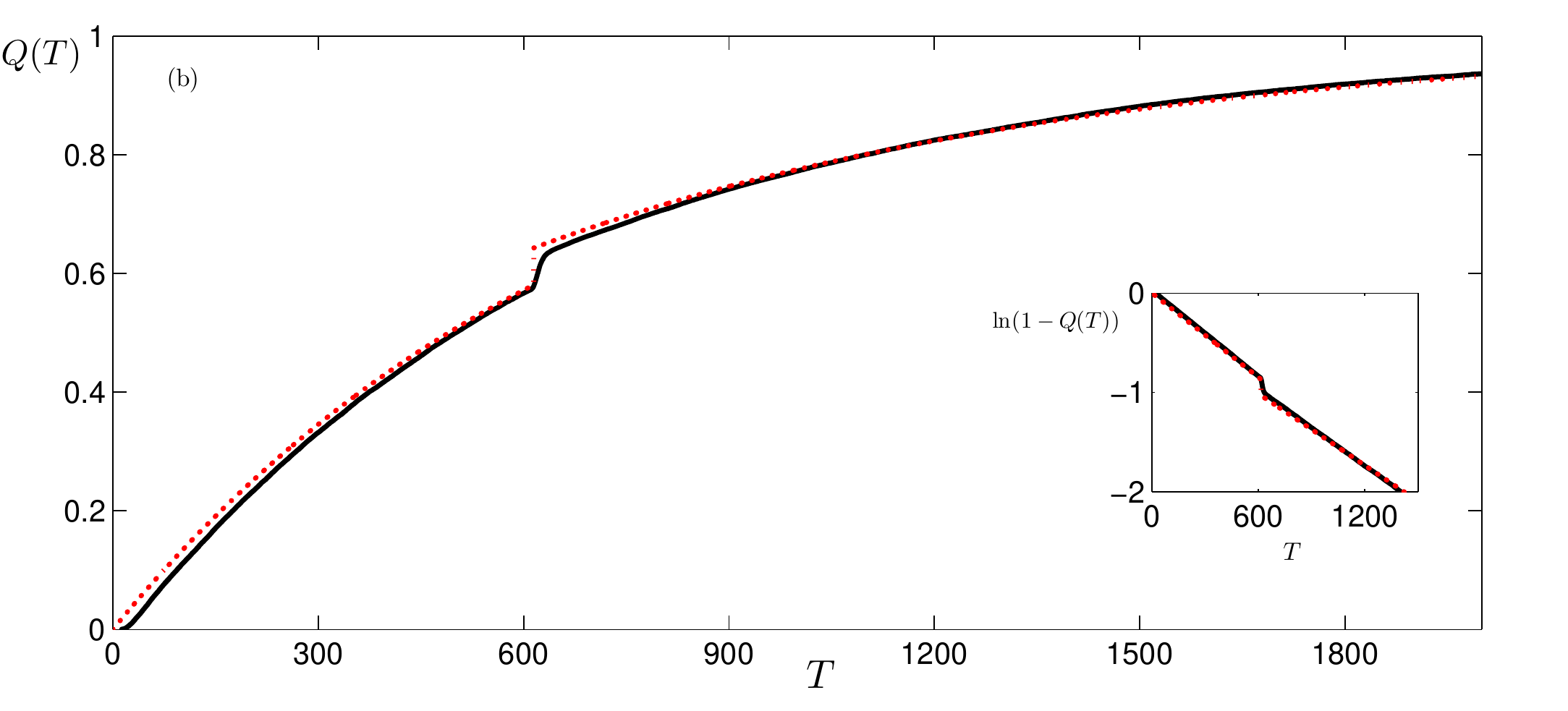}
	\caption{Cumulative ISI distribution $Q(T)$ vs $T$ for $n=2$ (panel (a)) 
	and $n=3$ (panel (b)) in a ring of unidirectional delay-coupled neurons.  
	The black lines are direct numerical simulations of Eq.~(\ref{Eq:model}), 
	where the values of parameters are chosen as: for $n=2$, $\hat{\tau}_{1}=100$, 
	$\hat{\tau}_{2}=200$ for $n=3$, $\hat{\tau}_{1}=100$, 
	$\hat{\tau}_{2}=200$, $\hat{\tau}_{3}=300$. 
	Values of $a$, $\epsilon$ and $D$ are chosen as $a=0.95$, $\epsilon=0.14$, $D=0.005$ f
	or both cases. The red dotted lines correspond to the 
	point process with Eq.~(\ref{Eq:cumulative}) for cumulative ISI, 
	where $\lambda=6.64\times10^{-4}$, $p=0.53$ are the same for 
	both cases. The effective time delays are $\tau_{1}=107, \tau_{2}=207$ 
	for $n=2$ case. $\tau_{1}=107, \tau_{2}=207, \tau_{3}=307$ for $n=3$ case.  
	The inset is a logarithmic scale version to validate the piecewise-linear spiking rate.}
	\label{fig:ISI}
\end{figure}

Based on the above description, and on the relationship between the cumulative 
ISI distribution $Q(T)$ and the survival function $S(T)$, which reads $Q(T)=1-S(T)$, 
the cumulative 
ISI distribution of neuron $i$ can be obtained as follows:
\begin{equation}  
	Q_{i}(T)=\begin{cases}
		1-e^{-\mu_{i} T}, & T<\tilde{\tau}, \\  
		1-(1-p_{i}p_{j})e^{-\mu_{i}\tilde{\tau}-(\lambda_{i}+\lambda_{j}p_{j})(T-\tilde{\tau})}, & T\geq \tilde{\tau},    
	\end{cases}\label{Eq:cumulative}
\end{equation} 
where $\tilde{\tau}=\tau_{i}+\tau_{j}$. We compare this expression with results
of numerical simulations in Fig.~\ref{fig:ISI}. The point process 
described by Eq.~(\ref{Eq:cumulative}) agrees well with direct simulations 
of Eq.~(\ref{Eq:model}), 
where we assumed the parameters of the two neurons to be the same, 
except for the time delays which are different.

\subsection{Correlations and spectra} 
In the following we derive the autocorrelation and the cross-correlation
functions of the spike trains, and the corresponding power spectrum 
and the cross-spectrum.
The autocorrelation function is defined via a joint probability 
to have a spike in unit $i$ within a small time interval $(t,t+\Delta t)$, 
and a spike in unit $j$ within
the time interval $(t+s,t+s+\Delta t)$, 
no matter whether or not there are any spikes between $t$ and $t+s$.
The joint probability of these events is defined as 
\begin{equation}
P_{ij}(t,t+\Delta t; t+s,t+s+\Delta t)=\begin{cases}
W_{i}(t,t+\Delta t)P_{ij}(t+s|t,\Delta t), &s\geq 0,\\
W_{j}(t,t+\Delta t)P_{ji}(t|t+s,\Delta t), &s<0.\end{cases}
\label{Eq:jointpro_ij}
\end{equation}
Here $W_{i}(t,t+\Delta t)=\mu_{i}\Delta t$ is the probability to observe a 
spike in neuron $i$ within the time interval $[t,t+\Delta t]$.
The quantity $P_{ij}(t+s|t,\Delta t)$ 
is the probability to induce a spike in 
neuron $j$ at time $t+s$, given a spike in neuron $i$ at time $t$.
\subsubsection{Correlations and spectra within one unit}
We first calculate the conditional probability~\eqref{Eq:jointpro_ij} 
for the same unit. The conditional probability to have one induced
spike is $p_ip_j$, this event happens with time shift $\tilde\tau$;
the conditional probability for the $k$-th induced spike is $(p_ip_j)^k$, it
happens with delay $k\tilde\tau$. Therefore,
\begin{equation}
\begin{aligned}
P_{ii}(t+s|t,\Delta t)&=\delta(s)\Delta t+p_{i}p_{j}\delta(s-\tilde{\tau})\Delta t+
\cdots+(p_{i}p_{j})^k\delta(s-k\tilde{\tau})\Delta t+\cdots\\&=
\sum\limits_{k=0}^{\infty}(p_{i}p_{j})^{k}\delta(s-k\tilde{\tau})\Delta t,\quad s \geq 0;\\
P_{ii}(t|t+s,\Delta t)&=P_{ii}(t-s|t,\Delta t), s<0,
\end{aligned}\label{Eq:condi_ii}
\end{equation}
where $\delta(\cdot)$ is the Dirac delta function.
Since the correlation function can be seen as the mean rate of 
the joint event, after substituting Eq.~(\ref{Eq:jointpro_ij}) 
and (\ref{Eq:condi_ii}), the auto-correlation function is
\begin{equation}
\begin{aligned}
C_{ii}(s)&=\langle (x_{i}(t) -\langle x_i\rangle)(x_{i}(t+s)-\langle x_i\rangle)\rangle\\&=
\frac{1}{T}\int_{0}^{T}dt\lim\limits_{\Delta t\rightarrow 0}\frac{P_{ii}(t,t+\Delta t; t+s,t+s+\Delta t)}{\Delta t^2}
=\mu_{i}\sum\limits_{n=-\infty}^{\infty}(p_{i}p_{j})^{|n|}\delta(s-n\tilde{\tau}),
\end{aligned}\label{Eq:correlation_ii}
\end{equation}
where we use the fact that $P_{ii}$ is $t$-independent, and have taken into account
that $\langle x_i\rangle=\mu_i$.

\begin{figure*}
	\centering
	\includegraphics[width=0.85\textwidth]{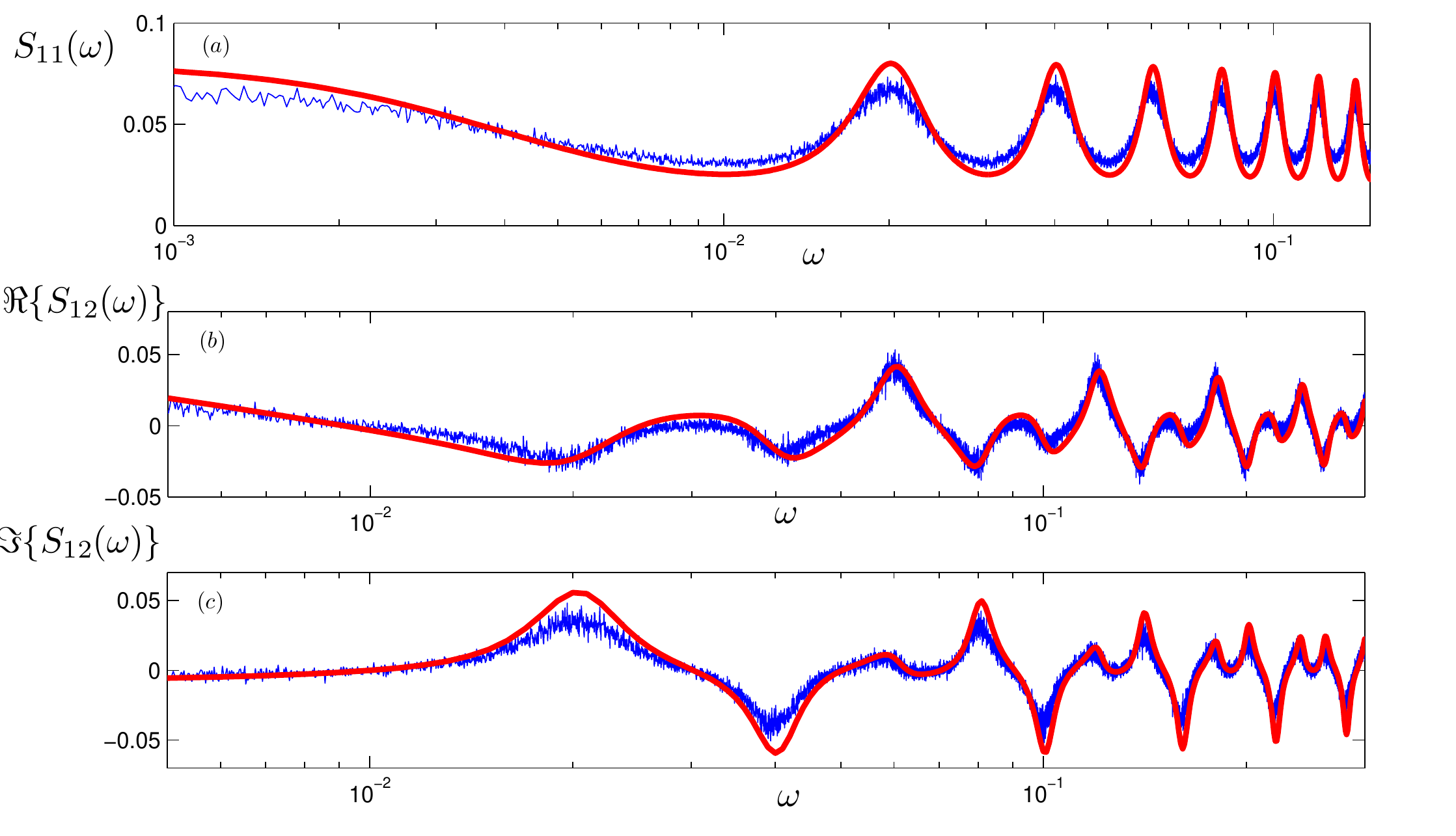}
	\includegraphics[width=0.85\textwidth]{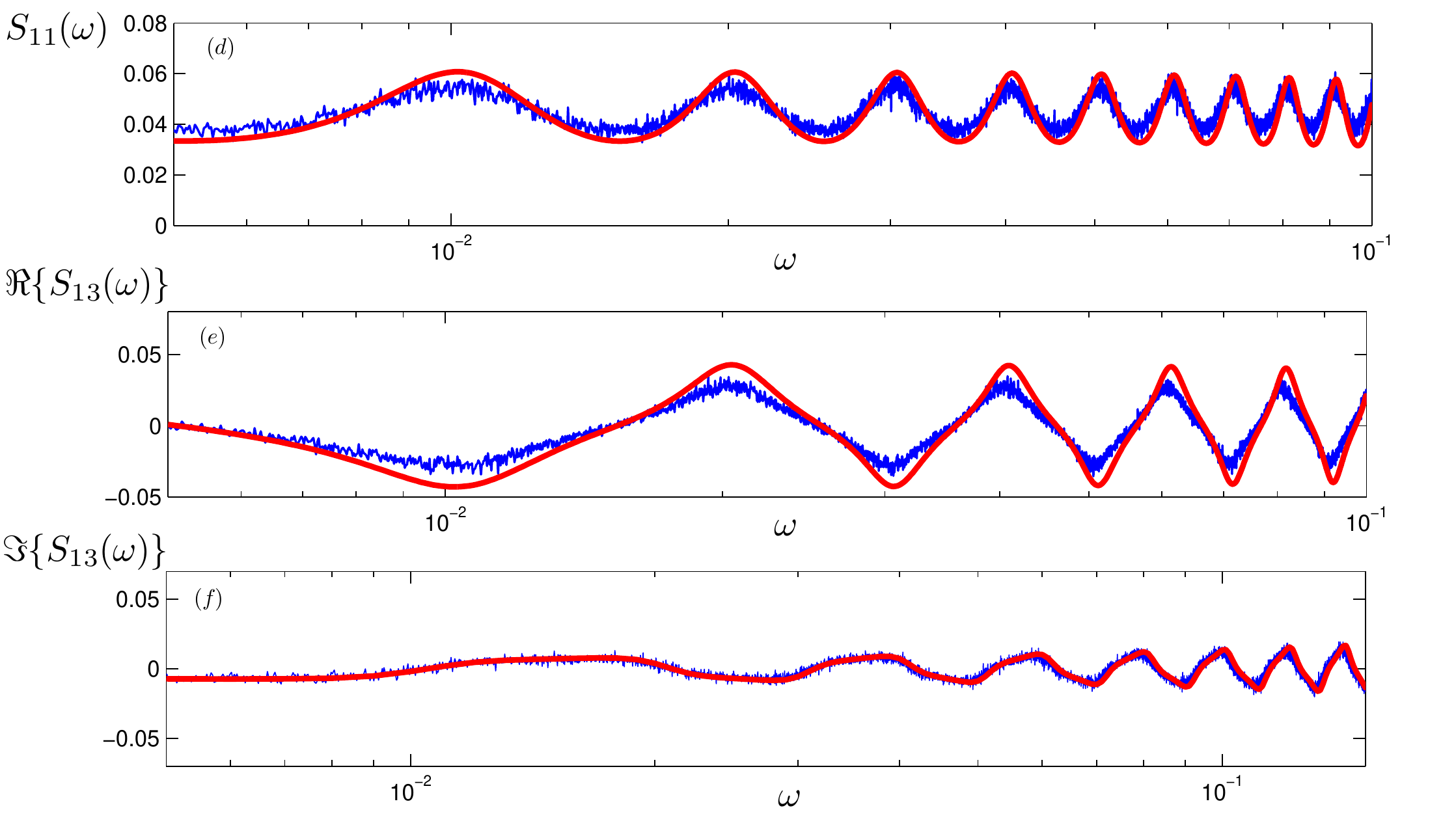}
	\caption{Panels (a)-(c) show the power spectral density of neuron 1, and
	the real part  and the imaginary part of the cross-spectral density $S_{12}$, 
	respectively, for two delay-coupled neurons. The blue lines are 
	from direct simulation of Eq.~(\ref{Eq:model}) and the red lines are 
	the analytical results from Eq.~(\ref{Eq:power_2}), Eq.~(\ref{Eq:crsp_2osci_Re}) 
	and Eq.~(\ref{Eq:crsp_2osci_Im}) for (a)-(c), respectively. Panels (d)-(f) show 
	the power spectral density, the real part  and the imaginary part of the 
	cross-spectral density $S_{13}$, respectively, for the ring of $n=3$ neurons. 
	The blue lines show numerical simulations of Eq.~(\ref{Eq:model}) with $n=3$. 
	The red lines are the analytical results from Eq.~(\ref{Eq:power_n}), 
	Eq.~(\ref{Eq:crsp_nosci_Re}) and Eq.~(\ref{Eq:crsp_nosci_Im}) for (d)-(f), respectively.
		The parameter values in the simulation and the analytical expressions are 
		chosen the same as in Fig.~\ref{fig:ISI}. 
		Noteworthy, all the power and corss-spectral density are multiplied by the power 
		spectral density of the shape function $S_{H}$, similar to Ref. \onlinecite{zheng2018delay}.}
	\label{fig: crsp_2&3osci}
\end{figure*}

Taking the Fourier transform of the correlation function, we obtain the power spectral density 
\begin{equation}
\begin{split}
S_{ii}(\omega)=\int\limits_{-\infty}^{\infty}C_{ii}(s)e^{-i\omega s}ds
=\frac{(\lambda_{i}+\lambda_{j}p_{j})(1+p_{i}p_{j})}{1+(p_{i}p_{j})^2-2p_{i}p_{j}
\cos\omega(\tau_{i}+\tau_{j})}. \label{Eq:power_2}
\end{split}
\end{equation}
The derivations above are based on the time series $x_{i}(t)$ represented as a sum of 
delta-peaks, i.e., $x_{i}(t)=\sum\limits_{i=1}^{N}\delta(t-t_{i})$. 
For a train of realistic spikes, the shape function can be straightforwardly 
taken into account as done in Ref.~\onlinecite{zheng2018delay}, namely the spectrum 
\eqref{Eq:power_2} should be just multiplied by the
squared amplitude of the Fourier transform of the pulse shape. For example, 
if observable \eqref{Eq:forc} is used,
 the spike train $x_{i}$ will be convoluted with the shape function $H(t)$. Hence, 
the power spectral density and the cross-spectral density in the following illustrations
will be multiplied 
by the spectral density of $H(t)$, which we denote as $S_{H}(\omega)$. For simplicity, 
in the formulas below we 
still use the delta-peak representation of $x_{t}$, while we 
multiply by $S_{H}(\omega)$ to compare 
with numerical correlations and spectra obtained by simulations 
of Eq.~(\ref{Eq:model}).
This comparison is shown in Fig.~\ref{fig: crsp_2&3osci}(a), 
the theoretical predictions
based on the point process analysis agree well with the results 
of direct simulation of Eq.~(\ref{Eq:model}).
\subsubsection{Cross-correlations and cross-spectra for two units}
The conditional probability of the joint event between the 
two neurons can be expressed similarly to 
formula~\eqref{Eq:condi_ii} above:
\begin{equation}
\begin{aligned}
P_{ij}(t+s|t,\Delta t)=&p_{i}\delta(s-\tau_{i})\Delta t+\cdots+p_{i}(p_{i}p_{j})^k\delta(s-k\tilde{\tau}-\tau_{i})\Delta t+\cdots
\\&=\sum\limits_{k=0}^{\infty}p_{i}(p_{i}p_{j})^{k}\delta(s- k\tilde{\tau}-\tau_i)\Delta t,\quad s \geq 0;\\
P_{ji}(t|t+s,\Delta t)=&p_{j}\delta(s+\tau_{j})\Delta t+\cdots+p_{j}(p_{i}p_{j})^k\delta(s+k\tilde{\tau}+\tau_{j})\Delta t+\cdots
\\&=\sum\limits_{k=0}^{\infty}p_{j}(p_{i}p_{j})^{k}\delta(s+k\tilde{\tau}+\tau_{j})\Delta t,\quad s<0.
\end{aligned} \label{Eq:condi_ij}
\end{equation}
This allows us to obtain the cross-correlation function of the two neurons, 
by substituting Eq.~(\ref{Eq:condi_ij}) into Eq.~(\ref{Eq:jointpro_ij}), 
leading to the Eq.~(\ref{Eq:correlation_ij}):
	\begin{equation}
	\begin{aligned}
	C_{ij}(s)=&\lim\limits_{\Delta t\rightarrow 0}\frac{P_{ij}(t,t+\Delta t; t+s,t+s+\Delta t)}{\Delta t^2}\\
	&=\mu_{i}\sum\limits_{n=0}^{\infty}p_{i}(p_{i}p_{j})^{n}\delta(s-n\tilde{\tau}-\tau_{i})+
	\mu_{j}\sum\limits_{n=0}^{\infty}p_{j}(p_{i}p_{j})^{n}\delta(s+n\tilde{\tau}+\tau_{j})\;.
	\end{aligned}\label{Eq:correlation_ij}
	\end{equation}

The cross-spectral density is the Fourier transform of 
the cross-correlation function:
\begin{equation}
S_{ij}(\omega)=\int\limits_{-\infty}^{\infty}C_{ij}(s) e^{-i\omega s}ds\\
=\mu_{i}p_{i}\frac{e^{-i\omega\tau_{i}}}{1-p_{i}p_{j}e^{-i\omega\tilde{\tau}}}+
\mu_{j}p_{j}\frac{e^{i\omega\tau_{j}}}{1-p_{i}p_{j}e^{i\omega\tilde{\tau}}}.
\label{Eq: crospec_2_general}
\end{equation}
It is instructive to present explicitly the real part
\begin{equation}
\Re\{{S_{ij}}\}=\frac{p_{i}(\mu_{i}-\mu_{j}p_{i}p_{j})\cos\omega\tau_{i}+p_{j}(\mu_{j}-\mu_{i}p_{i}p_{j})\cos\omega\tau_{j}}{1+(p_{i}p_{j})^2-2p_{i}p_{j}\cos\omega\tilde{\tau}}, \label{Eq:crsp_2osci_Re}
\end{equation}
and the imaginary part
\begin{equation}
\Im\{{S_{ij}}\}=\frac{p_{i}(\mu_{i}-\mu_{j}p_{i}p_{j})\sin\omega\tau_{i}-p_{j}(\mu_{j}-\mu_{i}p_{i}p_{j})\sin\omega\tau_{j}}{1+(p_{i}p_{j})^2-2p_{i}p_{j}\cos\omega\tilde{\tau}}. \label{Eq:crsp_2osci_Im}
\end{equation}
of the cross-spectrum.

Unlike the power spectral density described by a real-valued function 
(\ref{Eq:power_2}), the cross-spectral density is generally a 
complex-valued function. It is real-valued only when the two neurons are 
totally identical, i.e., $\lambda_{i}=\lambda_{j}=\lambda$, $p_{i}=p_{j}=p$, 
and $\tau_{1}=\tau_{2}=\tau$, resulting in a simple expression 
\begin{equation}
\begin{split}
S_{ij}(\omega)=\frac{2\lambda p(1+p)\cos\omega\tau}{1+p^4-2p^2\cos2\omega\tau},
\end{split} \label{Eq: crospec_2_identical}
\end{equation}
which is very similar to the power spectral density of a single unit~\eqref{Eq:power_2}.
We compare the theoretical cross-spectra with simulations
in Figs.~\ref{fig: crsp_2&3osci}b,c.

\subsubsection{Correlation and spectra of the total output from the network}
If we consider correlations and spectra from 
the viewpoint of the total network output, 
the cross-correlations between all the pulses should be calculated.
A joint probability could be defined as having a 
spike in any unit within a small time interval $(t,t+\Delta t)$, 
and a spike in any unit within
the time interval $(t+s,t+s+\Delta t)$, 
no matter whether or not there are any spikes between $t$ and $t+s$.
The joint probability $\hat{P}$ of these events is the sum of all contributions: 
\begin{equation}
\hat{P}(t,t+\Delta t; t+s,t+s+\Delta t)=\sum_{i=1}^{2}\sum_{j=1}^{2}P_{ij},
\end{equation}
where $P_{ij}$ is described by Eq.~(\ref{Eq:jointpro_ij}). 
Thus the correlation function is
\begin{equation}
\hat{C}(s)=\frac{1}{T}\int_{0}^{T}dt
\lim\limits_{\Delta t\rightarrow 0}
\frac{\hat{P}(t,t+\Delta t; t+s,t+s+\Delta t)}{\Delta t^2}=
\sum_{i=1}^{2}\sum_{j=1}^{2}C_{ij}
\end{equation}
where $C_{ij}$ is described by Eq.~(\ref{Eq:correlation_ii}) 
when $i=j$ and by Eq.~(\ref{Eq:correlation_ij}) when $i\neq j$. 
The spectral density of the total output , i.e. of the observable 
$X(t)=x_1(t)+x_2(t)$, is obtained as a
 Fourier transform of $\hat{C}(s)$, leading to
\begin{equation}
S_{X}(\omega)=\sum\limits_{i=1}^{2}\mu_{i}\frac{1-(p_ip_j)^2+
2p_i(\cos\omega\tau_i-p_ip_j\cos\omega\tau_j)}{1+(p_ip_j)^2-2p_ip_j
\cos\omega\tilde{\tau}}. \label{Eq:power_output_2}
\end{equation}
\begin{figure}
	\centering
	\includegraphics[width=\textwidth]{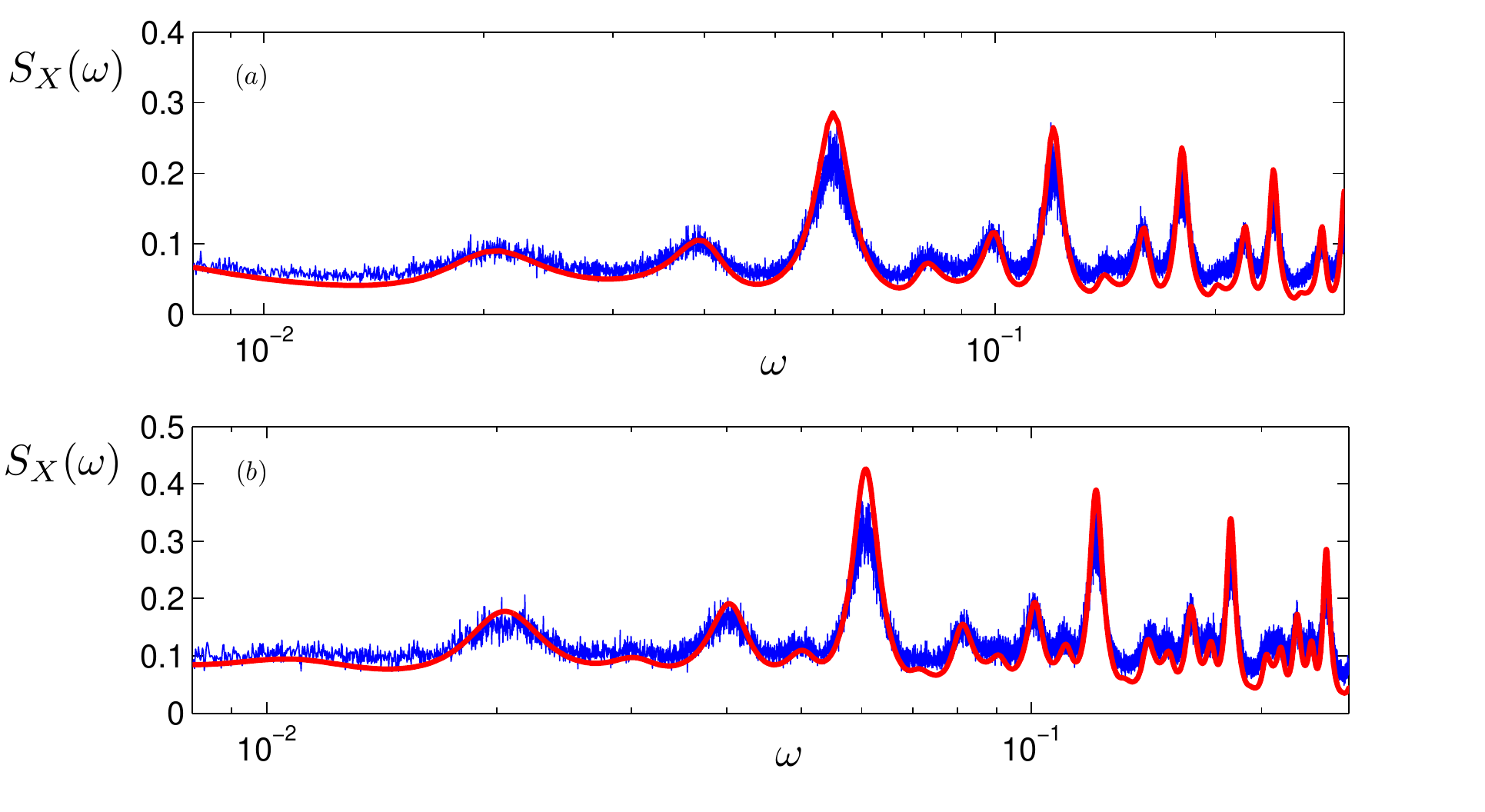}
	\caption{Power spectral density of the total output from the networks with $n=2$ 
	(panel (a)) and  $n=3$ (panel (b)). The blue 
	curves correspond to the simulation results and the red lines are the 
	theoretical expressions Eq.~(\ref{Eq:power_output_2}) for $n=2$ 
	and Eq.~(\ref{Eq:power_output_n}) for $n=3$. Values of parameters 
	are chosen the same as in Fig.~\ref{fig:ISI} and Fig.~\ref{fig: crsp_2&3osci}. }
	\label{fig:power_output}
\end{figure}
As is shown in Fig.~\ref{fig:power_output} (a), the theoretical spectra of the total 
output agrees well with simulation results.

\section{General network}
The case of many neurons with $n>2$ in the ring topology is a direct extension 
of the $n=2$ case as described above; thus the analysis follows the same steps, only
the expressions are more involved. 
First, we extend the cumulative ISI distribution for neuron $i$ in the ring as follows,
\begin{equation}  
Q_{i}(T)=\begin{cases}
1-e^{-\mu_{i} T}, & T<\tilde{T}, \\  
1-(1-\tilde{P})e^{-\mu_{i}\tilde{T}-\tilde{\mu}_{i}(T-\tilde{T})}, & T\geq \tilde{T}.    
\end{cases}       \label{Eq:cumu_n}
\end{equation} 
Here $\tilde{T}=\sum\limits_{i=1}^{n}\tau_{i}$ is the total round-trip delay time acround 
the ring, $\tilde{P}=\prod\limits_{j=1}^{n}p_{j}$ is the probability to have a completed
round trip around the ring, and $\tilde{\mu}_{i}$ is the spiking rate 
of all first spikes in bursts that include neuron $i$: 
\begin{equation}
\begin{aligned}
\tilde{\mu}_{i}=&\lambda_{i}+\lambda_{i-1}p_{i-1}+\lambda_{i-2}p_{i-2}p_{i-1}+\cdots+\lambda_{i-n+1}p_{i-1}\cdots p_{i-n+1}
\\&=\lambda_{i}+\sum\limits_{l=1}^{n-1}\lambda_{i-l}\prod\limits_{j=1}^{l}p_{i-j}.
\end{aligned}
\end{equation}
Here $\lambda_{i}$ is the rate of spontaneous spikes in neuron $i$ itself, 
$\lambda_{i-1}p_{i-1}$ is the rate of spontaneous spikes in neuron $i-1$ 
that induce also a spike in neuron $i$, and so on. 
Noteworthy, due to the ring structure of the coupling,
$\lambda_i$ is a periodic function, i.e., $\lambda_{i}=\lambda_{i+n}=\lambda_{i-n}$.
The total activity in the network is characterized by the rate $\mu_{i}$,
 to which the rates from all spikes (both leaders and followers) contribute.
 Thus, similarly to Eq.~\eqref{Eq:spikerate_2neurons}, we get
\begin{equation}
\mu_{i}=\lim\limits_{m\rightarrow\infty}\tilde{\mu}_{i}(1+\tilde{P}+\tilde{P}^2+\cdots+\tilde{P}^m)=
\frac{\tilde{\mu}_{i}}{1-\tilde{P}}.
\end{equation} 
Differently formulated, the expression above follows from the fact that a 
spike can have $m$ followers (in the same unit)
with probability $\tilde{P}^m(1-\tilde{P})$.

Using the same method as in the $n=2$ case described above, we obtain the power 
spectral density of neuron $i$ in the ring by Fourier transform of the 
correlation function (not presented):
\begin{equation}
\begin{aligned}
S_{ii}(\omega)=\frac{\tilde{\mu}_{i} (1+\tilde{P})}{1+\tilde{P}^2-2\tilde{P}\cos\omega\tilde{T}}. \label{Eq:power_n}
\end{aligned}
\end{equation}
The cross-spectral density of spike trains in neuron $i$ and neuron $j$ is:
\begin{equation}
S_{ij}(\omega)=\mu_{i}\bar{P}_{ij}\frac{e^{-i\omega T_{ij}}}{1-\tilde{P}e^{-i\omega\tilde{T}}}+\mu_{j}\bar{P}_{ji}\frac{e^{i\omega T_{ji}}}{1-\tilde{P}e^{i\omega\tilde{T}}},
\end{equation}
The real part of which is:
\begin{equation}
\Re\{{S_{ij}}\}=\frac{\bar{P}_{ij}(\mu_{i}-\mu_{j}\tilde{P})\cos\omega T_{ij}+\bar{P}_{ji}(\mu_{j}-\mu_{i}\tilde{P})\cos\omega T_{ji}}{1+\tilde{P}^2-2\tilde{P}\cos\omega\tilde{T}}, \label{Eq:crsp_nosci_Re}
\end{equation}
and the imaginary part of which is
\begin{equation}
\Im\{{S_{ij}}\}=\frac{\bar{P}_{ij}(\mu_{i}-\mu_{j}\tilde{P})\sin\omega T_{ij}-\bar{P}_{ji}(\mu_{j}-\mu_{i}\tilde{P})\sin\omega T_{ji}}{1+\tilde{P}^2-2\tilde{P}\cos\omega\tilde{T}}. \label{Eq:crsp_nosci_Im}
\end{equation}
Here $T_{ij}=\tau_{i}+\cdots+\tau_{j-1}$ is the delay time from neuron $i$ to neuron $j$ 
along the direction of the ring , i.e. clockwise as depicted in Fig.~\ref{fig:spike}, 
with probability $\bar{P}_{ij}=\prod\limits_{l=i}^{j-1}p_{l}$. 
Correspondingly, $T_{ji}=\tilde{T}-T_{ij}$ is the delay time from 
neuron $j$ to come to neuron $i$ with 
probability $\bar{P}_{ji}$ and $\bar{P}_{ij}\bar{P}_{ji}=\tilde{P}$.
The spectral density of total output, i.e. $X=\sum\limits_{i=1}^{n} x_{i}(t)$, 
from the network is
\begin{equation}
S_{X}(\omega)=\sum\limits_{i=1}^{n}\sum\limits_{j=i+1}^{i+n{\color{red}{-1}}}\mu_{i}\frac{1-\tilde{P}^2+2\bar{P}_{ij}(\cos\omega T_{ij}-\tilde{P}\cos\omega T_{ji})}{1+\tilde{P}^2-2\tilde{P}\cos\omega\tilde{T}}. \label{Eq:power_output_n}
\end{equation}
In the case that all the units are totally identical, i.e. $\lambda_{i}=\lambda$,
$p_{i}=p$ and $\tau_{i}=\tau (i=1, \cdots, n)$, $S_{X}(\omega)$ 
reduces to $S_{X}(\omega)=\frac{n\lambda(1+p)}{1+p^2-2\cos\omega\tau}$.

Generally, the model works for any network size $n$, but for simplicity we choose $n=3$ 
for  comparison with numerics. 
The cumulative ISI described by Eq.~(\ref{Eq:cumu_n}), spectra described 
by Eqs.~(\ref{Eq:power_n}), (\ref{Eq:crsp_nosci_Re}), (\ref{Eq:crsp_nosci_Im}) and (\ref{Eq:power_output_n}) agree well with 
direct simulation of Eq.~(\ref{Eq:model}), 
as shown in Fig.~\ref{fig:ISI}(b), Fig.~\ref{fig: crsp_2&3osci}(d)-(f) and Fig.~\ref{fig:power_output}(b), respectively. 
Noteworthy, similar to the case $n=2$,  the cross-spectrum $S_{ij}$ is generally a 
real-valued function only if $n$ is an even number and neurons $i$ and $j$ 
are symmetric, i.e., $|i-j|=n/2$. 

\begin{figure}
	\centering
	\includegraphics[width=\textwidth]{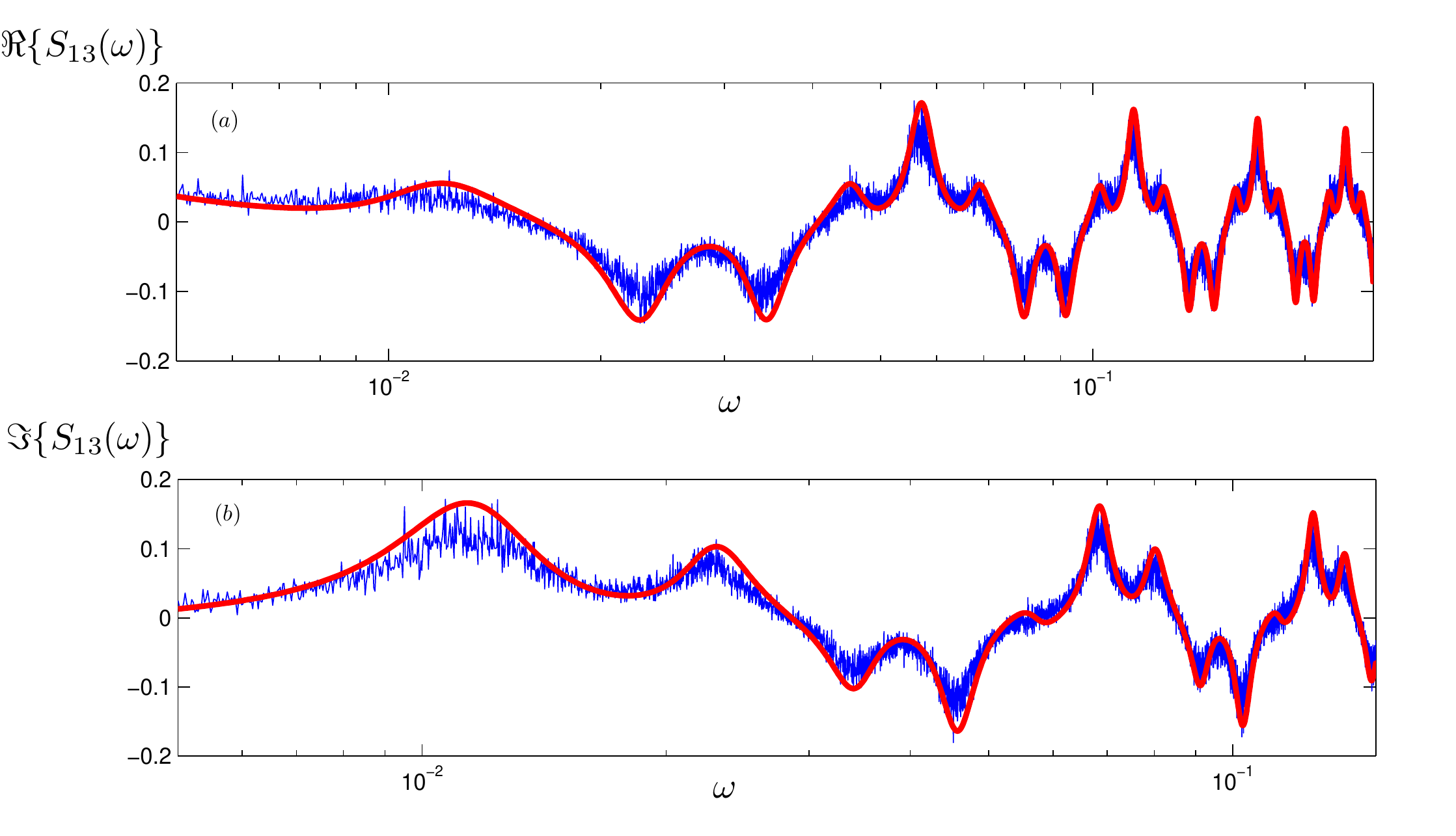}
	\includegraphics[width=\textwidth]{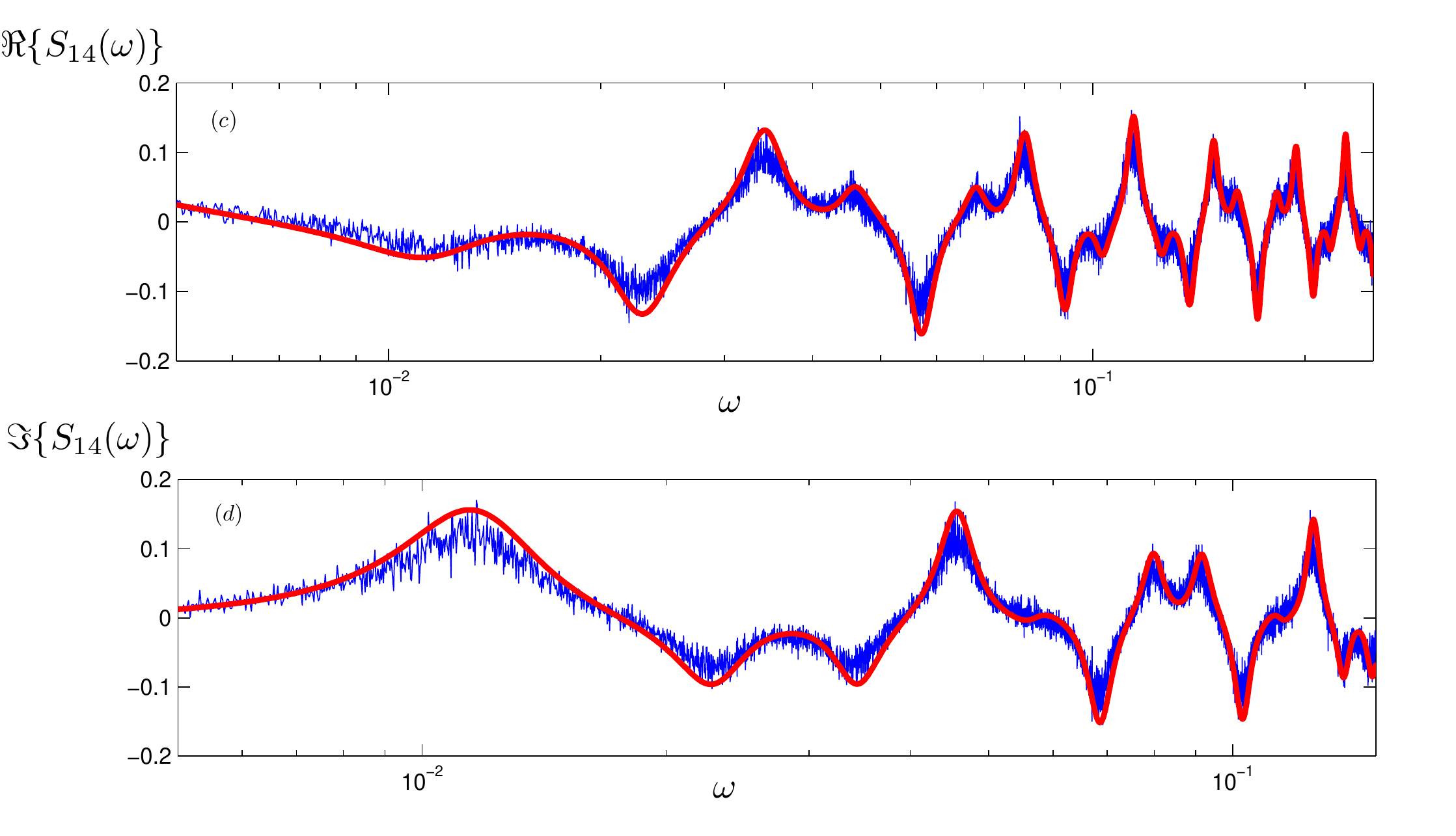}
	\caption{Cross-spectra between 1 and neuron 3 (panel (a) and panel (b) 
	for real and imaginary part respectively, 
	and between neuron 1 and 
	neuron 4 (panel (c) and (d) for real and imaginary part respectively). 	
	The blue lines show numerical simulations of Eq.~(\ref{Eq:model}) 
	where values of parameters are $a=0.95, D=0.005, \epsilon=0.2, 
	\tau_{i}=50 (i=1,\cdots, 10)$. The red lines are the analytical results from 
	Eq.~(\ref{Eq:crsp_nosci_Re}) and Eq.~(\ref{Eq:crsp_nosci_Im}), 
	where $\lambda_i=6.64\times10^{-4}$ is the same as described in 
	the $n=2$ and $n=3$ cases and $p_i=0.85 (i=1,\cdots, 10)$ is calculated 
	from Eq.~(\ref{Eq:p_simulation}).  }
	\label{fig:crpower_10neu}
\end{figure}

To further demestrate that our theory works for a
larger network, we choose $n=10$ and calculate the cross-spectra 
between neurons at different distances, e.g. between neurons 1 and 3, 
and between neurons 1 and 4. As shown in Fig.~\ref{fig:crpower_10neu}, 
the analytical results agree well with the simulations. 
Noteworthy, as $\epsilon$ goes larger, the duration of the delay-induced 
pulse becomes shorter, leading to a smaller empirical time shift $\bar{\tau}$. 
In the case depicted in Fig.~\ref{fig:crpower_10neu}, $\bar{\tau}\approx 5$ 
for $\epsilon=0.2$.

\section{Conclusion}
\label{sec:concl}

In conclusion, we investigated the stochastic bursting 
phenomenon in $n$ 
unidirectional delay-coupled noisy excitable systems. Under the condition 
of time-scale separation, an idealized version of coupled point 
processes with leader-follower relationship was formulated. Roughly speaking, 
occurrence of stochastic 
bursting is based on three ingredients: excitability of the system, 
excitatory coupling with a fixed time delay, and noise. Excitability combined with noise results 
in the spontaneous spikes with a constant spiking rate, which are leaders
of the bursts. A relatively weak coupling is not strong enough to induce a follower
deterministically, but it leads to an increased probability to have a follower, characterized
by the crucial parameter $p$. The leader with the followers form a burst, which
is rather coherent (because of the fixed time interval 
between the followers, nearly equal to the delay time), 
but has a random number of spikes in it.
 
 To characterize the stochastic bursting, the cumulative 
ISI distribution was derived; simulations demonstrated a good agreement
with the theoretical prediction. 
Furthermore, via the calculation the joint probability of the spikes, 
both the auto-correlation function 
of a single neuron spike train, the cross-correlation function of any pair of neurons in the 
unidirectional ring, and the auto-correlation function
of the total output from the network are derived analytically. Calculation of the
spectra and of the cross-spectra is then straightforward.
 Noteworthy, the model in the present paper 
not only shows an interesting coherent spiking pattern, but also
provides an alternative way to investigate the cross-spectrum of different 
neurons beyond the linear response theory 
(see, e.g., Refs.~\onlinecite{lindner2005theory,ostojic2009connectivity,trousdale2012impact,vullings2014spectra}, 
to name a few), which is widely used in the analysis of correlated neuronal networks.

Above we assumed, based on the time scale separation,
that the delay times are constants. A generalisation to the case of random
delay times is also possible and will be presented in details elsewhere;
here we discuss a simple version of this analysis. The essential point
where the fixed delays appear, is the representation of the correlation function
\eqref{Eq:correlation_ii} as a sum of delta-peaks at times $n\tilde\tau$. 
If we assume the delay times
to be independent Gaussian variables with mean value $\tilde\tau$ and standard deviation
$\kappa$, then one has to replace in~\eqref{Eq:correlation_ii}
delta-functions by Gaussian peaks 
$\delta(t-n\tilde\tau) \to (2n\pi\kappa^2)^{-1/2}\exp[-(t-n\tilde\tau)^2/(2n\kappa^2)]$. 
In the spectrum ~\eqref{Eq:power_2}, this correction corresponds to the
replacement $p_ip_j \to p_ip_j\exp[-n\omega^2\kappa^2/2]$. Around the main
frequency peaks (i.e. with small values of $n$), 
the effect of this correction is, as expected, small, due
to the time scale separation $\kappa \ll \tilde\tau$.

In this paper we restricted our attention
to a unidirectional coupling in the ring geometry, because here
overlapping of incoming spikes is not possible (or, better to say,
is very unprobable under the condition of the time-scale separation).
Such an overlap happens, e.g., in a network of delay-coupling neurons
demonstrating polychronization~\cite{Izhikevich06}; study of
stochastic bursting in such a setup is a subject of ongoing research.

\acknowledgments
C.Z. acknowledges the financial support from China Scholarship Council (CSC).
A.P. was supported by the Russian Science
Foundation (Grant No. 17-12-01534).

\end{document}